\documentclass[journal,10pt]{IEEEtran}
\ifCLASSINFOpdf
\else
   \usepackage[dvips]{graphicx}
\fi
\usepackage{url}
\usepackage{graphicx}
\usepackage{amsmath}
\usepackage{comment}
\usepackage[utf8]{inputenc}
\usepackage[english]{babel}
\usepackage{color}
\usepackage{framed}
\usepackage{amsmath,amssymb,amsfonts}

\usepackage{hyperref}

\usepackage{graphicx}
\usepackage{upgreek}
\usepackage{textcomp}
\usepackage{url}

\usepackage{cite}

\usepackage[dvipsnames]{xcolor}
\usepackage{amsmath}
\usepackage{amsthm}
\usepackage{dsfont}
\usepackage{thmtools}
\usepackage{amssymb}
\DeclareMathOperator*{\E}{\mathbb{E}}

\def\bnu{\boldsymbol{\nu}}

\def\btau{\boldsymbol{\tau}}

\def\bPhi{\boldsymbol{\Phi}}

\def\mbb{\mathbf{b}}
\def\mbc{\mathbf{c}}

\def\mbe{\mathbf{e}}

\def\mbp{\mathbf{p}}

\def\mbr{\mathbf{r}}

\def\mbv{\mathbf{v}}

\def\mbx{\mathbf{x}}
\def\mby{\mathbf{y}}
\def\mbz{\mathbf{z}}

\def\mbA{\mathbf{A}}

\def\mbI{\mathbf{I}}

\def\mbR{\mathbf{R}}

\def\mbW{\mathbf{W}}
\def\mbX{\mathbf{X}}

\def\calH{\mathcal{H}}

\def\bzero{\boldsymbol{0}}

\newtheorem{theorem}{Theorem}

\theoremstyle{definition}

\declaretheorem[style=definition,name=Remark,qed=$\blacksquare$]{remark}

\usepackage{bbold}
\usepackage{ifpdf}
\usepackage{times}
\usepackage{layout}
\usepackage{float}
\usepackage{afterpage}
\usepackage{amsmath}
\usepackage{amstext}
\usepackage{amssymb,bm}
\usepackage{latexsym}
\usepackage{color}
\usepackage{flexisym}
\usepackage{kantlipsum}
\usepackage{graphicx}
\usepackage{amsmath}
\usepackage{amsthm}
\usepackage{graphicx}
\usepackage[caption=false,font=footnotesize]{subfig}
\usepackage[inkscapeformat=png]{svg}
\usepackage[caption=false,font=footnotesize]{subfig}

\usepackage{booktabs}
\usepackage{multicol}

\usepackage{enumitem}

\usepackage[switch]{lineno}
\usepackage[named]{algo}
\usepackage[noend]{algpseudocode}
\usepackage{algorithm}
\usepackage{tikz}

\makeatletter
\newcommand*{\rom}[1]{\expandafter\@slowromancap\romannumeral #1@}
\makeatother

\begin{document}
\setlength{\abovedisplayskip}{3pt}
\setlength{\belowdisplayskip}{3pt}

\title{Deep Learning-Enabled One-Bit DoA Estimation}
\author{\IEEEauthorblockN{Farhang Yeganegi$^\star$, Arian Eamaz$^\star$, Tara Esmaeilbeig$^\star$, and Mojtaba Soltanalian}\\
\IEEEauthorblockA{{University of Illinois Chicago, Chicago, IL 60607, USA}}
\thanks{*~The first three authors contributed equally to this work.}
}
\markboth{
}
{Shell \MakeLowercase{\textit{et al.}}: Bare Demo of IEEEtran.cls for IEEE Journals}
\maketitle

\begin{abstract}
Unrolled deep neural networks have attracted significant attention for their success in various practical applications. In this paper, we explore an application of deep unrolling in the direction of arrival (DoA) estimation problem when  coarse quantization is applied to the measurements.
We present a  compressed sensing formulation  for DoA estimation from one-bit data in which estimating  target DoAs  requires recovering  a sparse signal from a limited number of severely quantized linear measurements. In particular, we exploit covariance recovery from one-bit dither samples. To recover the covariance of transmitted signal, the learned iterative shrinkage and thresholding algorithm (LISTA) is employed fed by one-bit data. We  demonstrate that the upper bound of estimation performance is governed by the recovery error of the transmitted signal covariance matrix.
Through numerical experiments, we demonstrate the proposed LISTA-based algorithm's capability in estimating target locations. The code 
employed in this study is  available online\footnote{\textcolor{blue}{\url{https://github.com/TaraEsmaeilbeig/one_bit_DoA_estimation.git}.}}.

\end{abstract}


\setlength{\abovedisplayskip}{3pt}
\setlength{\belowdisplayskip}{3pt}

\section{Introduction}
\label{intro}
The Direction of Arrival (DoA) estimation problem holds paramount significance in array processing, finding applications across radar, sonar, and wireless communications domains~\cite{Barthelme2021cov}. 
The conventional techniques for DoA estimation rely on the premise that analog array measurements are digitally represented with a substantial number of bits per sample, thereby allowing for the disregarding of resulting quantization errors. However, the costs of production and energy consumption associated with analog-to-digital converters (ADCs) increase dramatically as the number of quantization bits and sampling rate rise. This challenge is particularly pronounced in systems requiring multiple ADCs, such as large array receivers \cite{ho2019antithetic}. One immediate solution to address these challenges is the adoption of fewer bits for sampling.

\begin{figure}[t]
\centering
\input{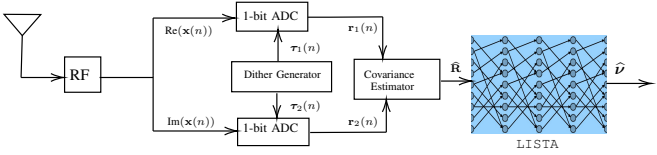}
\caption{A simplified illustration of the proposed system for DoA estimation from one-bit dithered data.} 
\label{fig_1}
\end{figure}

In recent years, there has been a growing emphasis on the design of receivers equipped with low-complexity \textit{one-bit ADCs} to satisfy the demands for wide signal bandwidth and low cost/power. \emph{One-bit quantization} represents a coarse scenario of quantization, wherein ADCs solely compare signals with predefined threshold levels, resulting in binary ($\pm1$) outputs. This approach enables signal processing equipment to sample at significantly higher rates while maintaining lower costs and energy consumption compared to conventional ADCs \cite{instrumentsanalog,mezghani2018blind,eamaz2021modified,10258362}. As a result, one-bit ADCs  are substantially  recognized in the field of DoA estimation \cite{liu2017one,sedighi2021performance,Huang2019music,YuDoA2016,zhou2020direction}. 
However, all of these efforts have relied on ditherless coarse quantization, resulting in a unidirectional estimation scenario where only the normalized version of the input covariance matrix can be recovered, leading to a significant loss of information. Introducing a random dither before quantization, as  shown in Fig.~\ref{fig_1}, demonstrates a significant enhancement in covariance recovery performance from quantized data. This enhancement is  investigated in the literature for various dithering schemes, including Gaussian and uniform dithering \cite{eamaz2022covariance,eamaz2023covariance,eamaz2021modified,dirksen2022covariance,10258362,dirksen2024tuning,xiao2023covariance}.

In this paper, we approach DoA estimation, as a sparse recovery problem, by means of  deep unrolling based compressed sensing solvers to recover the target locations. 
In recent years, researchers in the field of compressive sensing have endeavored to integrate \emph{model-driven} algorithms with \emph{data-driven} approaches to address sparse linear inverse problems, which involve recovering a sparse signal from a limited number of noisy linear measurements. These methods typically involve unfolding a well-established iterative recovery algorithm, such as iterative shrinkage and thresholding algorithm (ISTA), to construct a neural network with trainable variables. Subsequently, these trainable variables are optimized by means of optimization techniques such as  stochastic gradient descent algorithms.

The authors of \cite{gregor2010learning} introduced the learned ISTA (LISTA) framework, which incorporates learnable threshold variables for the shrinkage function. Such approaches offer the advantage of learning more realistic signal priors from the training data while retaining the interpretability of model-driven methods. Additionally, by explicitly accounting for the  underlying model, LISTA and similar methods can utilize smaller or shallower networks with fewer parameters compared to purely \emph{data-driven} counterparts \cite{chen2018theoretical}. This characteristic brings several benefits, including reduced demand for training data, mitigated risk of overfitting, and implementations with significantly reduced memory requirements \cite{monga2021algorithm, esmaeilbeig2023deep,hu2023ihtinspired,zheng2024antenna}. 

Authors of~\cite{xiao2020deepfpc} formulate the DoA estimation problem within the framework of compressed sensing, specifically with regard to the data matrix. The proposed algorithm is constructed by unrolling iterations of the fixed-point continuation algorithm, originally devised for addressing the one-bit compressed sensing problem. The algorithm's iterations are adapted to a deep unrolling approach. However, the work presented in\cite{xiao2020deepfpc} has two primary limitations. Firstly, using a ditherless scheme is a drawback. Additionally, the absence of any theoretical guarantees further diminishes the robustness of their approach.

In contrast, our paper not only surpasses~\cite{xiao2020deepfpc} by offering rigorous theoretical guarantees but also redefines the DoA estimation problem as a compressed sensing problem with respect to the input covariance matrix.

An additional related work is~\cite{khobahi2020model} in which the iterative process of the hard-thresholding algorithm has been unrolled  for one-bit compressed sensing with dithering. This study, introduces the dithers as learnable parameters, and in addition to storing quantized data, it requires storing  of dither values to be  used later in reconstruction. Moreover, they did not provide any  theoretical guarantees. In contrast, our approach, benefiting from uniform dithering,  eliminates the need to store dither values for the reconstruction phase and has theoretical guarantees for  reconstruction performance.

In the numerical results, we demonstrate the successful recovery of target locations using LISTA trained by one-bit data. Furthermore, we theoretically provide an upper bound for the recovery performance of the proposed algorithm.

\underline{\emph{Notation}}: Throughout this paper, we use bold lowercase and bold uppercase letters for vectors and matrices, respectively. We represent a vector $\mathbf{x}$ and a matrix $\mbX$ in terms of their elements as $\mathbf{x}=[x_{i}]$ and $\mathbf{X}=[\mbX]_{i,j}$, respectively. The sets of real and complex numbers are represented by $\mathbb{R}$ and $\mathbb{C}$, respectively. The mathematical
expectation is  returned by $\E\{\cdot\}$. The operators $(\cdot)^{\top}$, $(\cdot)^{\star}$, and $(\cdot)^{\mathrm{H}}$ denote the vector/matrix transpose, conjugate, and hermitian, respectively. The operator $\operatorname{diag}\left\{\mathbf{b}\right\}$ with $\mbb=[b_i]$ denotes a diagonal matrix with $\{b_{i}\}$ as its diagonal elements. The Khatri-Rao product is denoted by $*$.
The notation $x \sim \mathcal{U}_{[a,b]}$ means a random variable drawn from the uniform distribution over the interval $[a,b]$. The set $[n]$ is defined as $[n]=\left\{1,\cdots,n\right\}$. The $\ell_{p}$-norm of a vector $\mathbf{x}$ is $\|\mathbf{x}\|_{p}=\left(\sum_{i}x^{p}_{i}\right)^{1/p}$. The Frobenius norm of a matrix $\mathbf{X}\in \mathbb{C}^{n_1\times n_2}$ is defined as $\|\mathbf{X}\|_{\mathrm{F}}=\sqrt{\sum^{n_1}_{r=1}\sum^{n_2}_{s=1}\left|[\mbX]_{r,s}\right|^{2}}$. We also define $\|\mbX\|_{\mathrm{max}}=\sup_{i,j}|[\mbX]_{i,j}|$. The vectorized form of a matrix $\mbX$ is written as $\operatorname{vec}(\mbX)$. If there exists a $c>0$ such that $a\leq c b$ (resp. $a\geq c b$) for two quantities $a$ and $b$, we have $a \lesssim b$ (resp.  $a \gtrsim b$). For $a\in\mathbb{C}$, $\operatorname{Re}(a)$ and $\operatorname{Im}(a)$ denote the real and imaginary parts of $a$, respectively. For $a\in\mathbb{C}$, the sign operator $\operatorname{sgn}(\cdot)$ is denoted as $\operatorname{sgn}(a)=\operatorname{sgn}\left(\operatorname{Re}\left(a\right)\right)+\mathrm{j}\operatorname{sgn}\left(\operatorname{Im}\left(a\right)\right)$. The bounded set $\mathcal{X}$ is defined as $\mathcal{X}(B, s) =\left\{\mbx\mid |x_i|\leq B, \forall i,\left\|\mbx\right\|_0 \leq s\right\}$. 
\section{Signal Model}\label{sec:2}

We assume a uniform linear array (ULA) with $M$ sensors is impinged by  $K$ narrowband sources.
The sensor spacing is set to $d$ and $\lambda$ denotes the wavelength. Denote the  steering matrix by  $\mbA=\left[\begin{array}{llll}\mathbf{a}\left(\theta_1\right) & \mathbf{a}\left(\theta_2\right) & \cdots & \mathbf{a}\left(\theta_K\right)\end{array}\right]^{\top}\in\mathbb{C}^{M\times K}$, where $\left[\mbA\right]_{m,k}=e^{-\mathrm{j}\frac{2\pi \left(m-1\right)d \sin{\theta_k}}{\lambda}}$. Denote by $s_k(n)$ the transmitted data of the $k$-th source
at the $n$-th snapshot. The array measurement at the $n$-th snapshot is given by 
\begin{equation}
\label{eq1}
\begin{aligned}
\mathbf{x}(n) & =\left[\begin{array}{llll}
x_1(n) & x_2(n) & \cdots & x_M(n)
\end{array}\right]^{\top}\in\mathbb{C}^{M} \\
& =\sum_{k=1}^K \mathbf{a}\left(\theta_k\right) s_k(n)+\mbv(n) \\
& =\mathbf{A s}(n)+\mbv(n),\quad\text{for}~ n\in [N],
\end{aligned}
\end{equation}
where $\mbv(n)$ is a noise following $\mathcal{N}\left(0,\sigma^2\mbI\right)$. The DoA of the $k$-th source does not necessarily belong to the  overcomplete  DoA set $\{\bar{\theta}_1,\cdots,\bar{\theta}_L\}$, therefore
 the transmitted signal is zero-padded to achieve an overcomplete formulation, as follows:
\begin{equation}
p_l(n)=\left\{\begin{array}{cc}
s_k(n), & \left|\bar{\theta}_l-\theta_k\right| \leq \frac{\Delta \theta}{2}, \\
0, & \text {otherwise},
\end{array}\right. \quad l\in[L],~n\in [N],
\end{equation}
where
$\Delta \theta$ denotes  the interval of the  overcomplete  DoA set~\cite{su2023real,Wu2022DoA}. 
Consequently, the overcomplete form of~\eqref{eq1} is
\begin{equation}
\label{eq2}
\begin{aligned}
\mathbf{x}(n) &
=\sum_{l=1}^L \mathbf{a}\left(\bar{\theta}_l\right) p_l(n)+\mbv(n) \\
&=\widehat{\mbA}\mbp(n)+\mbv(n),\quad\text{for}~ n\in [N].
\end{aligned}
\end{equation}
The covariance matrix of $\mbx(n)$ is given by
\begin{equation}
\label{eq3}
\begin{aligned}
\mathbf{R} & =\E\left\{\mathbf{x}(n) \mathbf{x}^{\mathrm{H}}(n)\right\}\in\mathbb{C}^{M\times M} \\
& =\widehat{\mbA}\operatorname{diag}\left(\left[\begin{array}{llll}
\nu_1 & \nu_2 & \cdots & \nu_L
\end{array}\right]^{\top}\right) \widehat{\mbA}^{\mathrm{H}}+\sigma^2 \mbI_M,
\end{aligned}
\end{equation}
where $\{\nu_l\}^{L}_{l=1}$ are variances of $\{p_l(n)\}_{l=1}^L$. The covariance equation \eqref{eq3} can be linearized as
\begin{equation}
\label{eq4}
\begin{aligned}
& \mathbf{y}=\operatorname{vec}(\mathbf{R})\in\mathbb{C}^{M^2} \\
& =\left(\begin{array}{llll}
\widehat{\mbA}^{\star} * \widehat{\mbA}
\end{array}\right)\left[\begin{array}{llll}
\nu_1 & \nu_2 & \cdots & \nu_L
\end{array}\right]^{\top} \\
& +\sigma^2\left[\begin{array}{llll}
\mbe_1^{\top} & \mbe_2^{\top} & \cdots & \mbe_M^{\top}
\end{array}\right]^{\top},
\end{aligned}
\end{equation}
where $\mbe^{\top}_m$ is the $m$-th column of identity matrix. 
To enable better processing of measurements in our compressed sensing solver, we concatenate the real and complex values as follows:
\begin{equation}
\begin{aligned}
\label{q1}
\left[\begin{array}{l}
\operatorname{Re}(\mathbf{y}) \\
\operatorname{Im}(\mathbf{y})
\end{array}\right] &=  \left[\begin{array}{l}
\operatorname{Re}\left(\widehat{\mbA}^{\star} * \widehat{\mbA}\right)\\
\operatorname{Im}\left(\widehat{\mbA}^{\star} * \widehat{\mbA}\right)
\end{array}\right]\boldsymbol{\nu} \\&+\left[\begin{array}{cccc}
\operatorname{Re}\left(\sigma^{ 2} \left[\mbe_1^{\top}\right.\right. & \mbe_2^{\top} & \cdots & \left.\left.\mbe_M^{\top}\right]^{\top}\right) \\
&\bzero_{_{M^2 \times 1}}&&
\end{array}\right],
\end{aligned}    
\end{equation}
where $\boldsymbol{\nu}=\left[\begin{array}{llll}
\nu_1 & \nu_2 & \cdots & \nu_L
\end{array}\right]^{\top}\in\mathbb{R}^{L}$. In the next section, we will apply  coarse quantization on the measurements $\{\mbx(n)\}_{n=1}^{N}$.
\section{LISTA for One-Bit DoA estimation}
\label{sec:3}
We implement the one-bit quantization on complex measurements 
by employing two complex random dithering sequences $\{\boldsymbol{\uptau}_{1}(n),\boldsymbol{\uptau}_{2}(n)\}_{n=1}^{N}$, where  real and imaginary parts of each follows
the uniform distribution $\mathcal{U}_{\left[-T, T\right]}$ as 
\begin{equation}
\label{one-bit}
\left\{\begin{array}{cc}
\mbr_1(n)=\operatorname{sgn}\left(\mbx(n)+\boldsymbol{\uptau}_1(n)\right), \\
\mbr_2(n)=\operatorname{sgn}\left(\mbx(n)+\boldsymbol{\uptau}_2(n)\right),
\end{array}\right. \quad n\in [N].
\end{equation}
In our proposed scheme for DoA estimation, Illustrated in Fig.~\ref{fig_1}, one only has access to one-bit samples of the signal rather than high-resolution data. To estimate the covariance matrix of the signal from the one-bit data, i.e. estimate
$\mby$, we employ the following sample covariance estimation:
\begin{equation}
\label{cov_est}
\widehat{\mbR} = \frac{\widehat{\mbR}_1+\widehat{\mbR}^{\mathrm{H}}_1}{2},
\end{equation}
where $\widehat{\mbR}_1$ is
\begin{equation}
\widehat{\mbR}_1=\frac{T^2}{N}\sum^{N}_{n=1} \mbr_1(n)\mbr^{\mathrm{H}}_2(n).
\end{equation}
It has been comprehensively presented that this sample covariance estimation from the one-bit data $\widehat{\mbR}$ is an unbiased estimator, if the dynamic range of signal is restricted by the scale parameter of uniform dither, i.e., $\sup_{m,n} x_m(n)< T$ \cite{dirksen2022covariance}.   
We define $\bar{\mby}=\operatorname{vec}\left(\widehat{\mbR}\right)$, then combining the covariance estimate in \eqref{cov_est} with the model in \eqref{q1} leads to
\begin{equation}\label{q2}
\mbb=\boldsymbol{\Phi}\boldsymbol{\nu}+\mbz,
\end{equation}
where $\boldsymbol{\Phi}=\left[\begin{array}{l}
\operatorname{Re}\left(\widehat{\mbA}^{\star} * \widehat{\mbA}\right)\\
\operatorname{Im}\left(\widehat{\mbA}^{\star} * \widehat{\mbA}\right)
\end{array}\right]\in\mathbb{R}^{2M^2\times L}$, and $\mbb=\left[\begin{array}{l}
\operatorname{Re}(\bar{\mby}) \\
\operatorname{Im}(\bar{\mby})
\end{array}\right]\in\mathbb{C}^{2 M^2}$.  Then to recover the sparse unknown parameter $\boldsymbol{\nu}$ in \eqref{q2}, we solve 
\begin{equation}\label{eq:11}
\underset{\bnu}{\text{minimize}} \quad \frac{1}{2}\|b-\bPhi \bnu\|_2^2+\lambda\|\bnu\|_1.
\end{equation}
Inspired by ISTA, we utilize a model-based deep neural network called LISTA to tackle problem \eqref{eq:11}, as shown in Fig.~\ref{fig_1}.
Define $\operatorname{soft}_{\eta}(x)=\operatorname{sgn}(x)\max\left(0,|x|-\eta\right)$ as the soft thresholding operator.
Then, the $i$-th layer of  LISTA  neural network takes the form
\begin{equation}
\label{q3}
\boldsymbol{\nu}^{(i+1)}=\operatorname{soft}_{\eta^{(i)}}\left(\mbW_1^{(i)}\mbc+\mbW_2^{(i)}\boldsymbol{\nu}^{(i)}\right),~ i\in\{0,\cdots,I-1\},
\end{equation}
where the parameters $\mathcal{H}=\left\{\left(\mbW_1^{(i)},\mbW_2^{(i)},\eta^{(i)}\right)\right\}_{i=0}^{I-1}$ are subject to learning and $\mbc=\mbb-\mbz$. Note that LISTA is treated as a specially structured neural network and trained over a given training dataset $\left\{\left(\boldsymbol{\nu}_j,\mbc_j\right)\right\}_{j=1}^{S}$ with $S$ representing the size of the training dataset. The training  process is modeled as
\begin{align}\label{q4}
\underset{\mathcal{H}}{\textrm{minimize}}\quad\mathbb{E}_{\boldsymbol{\nu},\mbc}\left\|\boldsymbol{\nu}^{(I)}\left(\mathcal{H},\mbc,\boldsymbol{\nu}^{(0)}\right)-\boldsymbol{\nu}\right\|_2^2.
\end{align}
The following  remark provides the necessary conditions for convergence of LISTA:
\begin{remark}\cite[Theorem~1]{chen2018theoretical}\label{rm_1}
The necessary conditions for LISTA to capture the desired parameter $\boldsymbol{\nu}$, i.e. $\boldsymbol{\nu}^{(i)}\rightarrow\boldsymbol{\nu}$ as $i\rightarrow\infty$, are
\begin{align}\label{q5}
& \mbW_2^{(i)}-\left(\mbI-\mbW_1^{(i)}\boldsymbol{\Phi}\right) \rightarrow \mathbf{0}, ~\text {as} ~ i \rightarrow \infty, \\
& \eta^{(i)} \rightarrow 0, ~ \text {as} ~ i \rightarrow \infty.
\end{align}
\end{remark}
Under conditions of Remark~1, layer $i$ of LISTA in \eqref{q3} reduces to
\begin{equation}
\label{q6}
\boldsymbol{\nu}^{(i+1)}=\operatorname{soft}_{\eta^{(i)}}\left(\boldsymbol{\nu}^{(i)}+\left(\mbW^{(i)}\right)^{\top}\left(\mbc-\boldsymbol{\Phi}\boldsymbol{\nu}^{(i)}\right)\right),
\end{equation}
where $i\in\{0,\cdots,I-1\}$. Interestingly, in the update process in \eqref{q6}, the learnable parameters reduce to $\mathcal{H}=\left\{\left(\mbW^{(i)},\eta^{(i)}\right)\right\}_{i=0}^{I-1}$. Algorithm~\ref{alg:LISTA-DoA} summarizes the implementation steps  for training  LISTA under  quantized measurements.
\begin{algorithm}[t]
\caption{Training of LISTA for one-bit DoA estimation}
\label{alg:LISTA-DoA}
\begin{algorithmic}[1]
\Statex\textbf{Input} The one-bit data $\{\mbr_{1}(n),\mbr_{2}(n)\}_{n=1}^{N}$, the scale parameter of uniform dithers $T$.
\Statex\textbf{Output} $\calH$.
\State $\widehat{\mbR}_1\gets\frac{T^2}{N}\sum^{N}_{n=1} \mbr_1(n)\mbr^{\mathrm{H}}_2(n)$.
\State $\widehat{\mbR}\gets\frac{\widehat{\mbR}_1+\widehat{\mbR}^{\mathrm{H}}_1}{2}$.
\State Form the model in \eqref{q2}, create the training dataset $\left\{\left(\boldsymbol{\nu}_j,\mbc_j\right)\right\}_{j=1}^{S}$, with $\mbc_j=\mbb_j-\mbz_j$.
\State Set $i$-th layer of LISTA according to~\eqref{q6}  and feed  LISTA with training dataset to obtain the  optimized set of parameters $\calH$ through backpropagation.
\Statex\textbf{Return} $\mathcal{H}$.
\end{algorithmic}
\end{algorithm}
Before presenting our main convergence result, the subsequent theorem demonstrates the feasibility of precise covariance estimation from observed one-bit data with high probability:
\begin{theorem}
\label{thr_2}
Let $\mbx \in \mathbb{C}^M$ with covariance matrix $\mathbb{E}\left[\mbx\mbx^{\mathbf{H}}\right]=\mbR$ and $J$-subgaussian coordinates. Let $\mbx(1), \ldots, \mbx(N) \stackrel{\text { i.i.d. }}{\sim} \mbx$. Then, there exists a constant $c>0$ which only depends on $J$, such that the covariance estimator $\widehat{\mbR}$ fulfills, for any $t \geq 0$, with a probability at least $1-8 e^{-c N t}$
\begin{equation}
\left\|\widehat{\mbR}-\mbR\right\|_{\operatorname{max}}\lesssim T^2 \sqrt{\frac{\log (M)+t}{N}},
\end{equation}
and
\begin{equation}
\left\|\widehat{\mbR}-\mbR\right\|_{\mathrm{F}} \lesssim T^2 \sqrt{\frac{M^2(\log (M)+t)}{N}}.
\end{equation} 
\end{theorem}
\begin{IEEEproof}
See {\cite[Theorem~1]{10258362}}.
\end{IEEEproof}
The following theorem presents the convergence of LISTA with one-bit data.
\begin{theorem}
\label{thr_1}
Let $\{\mbx(n)\}_{n=1}^{N}$ be i.i.d $J$-subgaussian random variables. Assume that the scale parameter of uniform dithers $T$ satisfies the dynamic range condition i.e. $\sup_{m,n} x_m(n)< T$. Given the parameters $\left\{\left(\mbW^{(i)},\eta^{(i)}\right)\right\}_{i=0}^{\infty}$ and $\boldsymbol{\nu}^{(0)}=0$, let $\left\{\boldsymbol{\nu}^{(i)}\right\}^{\infty}_{i=1}$ be generated by \eqref{q6}. If $\boldsymbol{\nu}\in\mathcal{X}\left(B,s\right)$ for a sufficiently small $s$, then there exists a sequence of parameters $\left\{\left(\mbW^{(i)},\eta^{(i)}\right)\right\}$ such that, for all $\boldsymbol{\nu}\in\mathcal{X}\left(B,s\right)$ and constants $c_1,c_2,C,t\geq 0$, with probability exceeding $1-8e^{-c_1Nt}$, we have
\begin{equation}
\label{bound}
\left\|\boldsymbol{\nu}^{(i)}-\boldsymbol{\nu}\right\|_2\leq s B e^{-c_2 i}+ CT^2M^2\sqrt{\frac{ \log(M)+t}{N}}.
\end{equation}
\end{theorem}
\begin{IEEEproof}
As shown in \cite[Theorem~2]{chen2018theoretical}, the convergence rate of LISTA for noisy measurements takes the form
\begin{equation}
\label{q7}
\left\|\boldsymbol{\nu}^{(i)}-\boldsymbol{\nu}\right\|_2\leq s B e^{-c_2 i}+C\sigma^{\prime},
\end{equation}
where in our work
\begin{equation}
\label{q8}
\left\|\operatorname{vec}\left(\mbR\right)-\operatorname{vec}\left(\widehat{\mbR}\right)\right\|_{1}\leq\sigma^{\prime}.
\end{equation}
Based on Theorem~\ref{thr_2}, the left-hand side of \eqref{q8} can be upper bounded by
\begin{equation}
\label{q9}
\left\|\operatorname{vec}\left(\mbR\right)-\operatorname{vec}\left(\widehat{\mbR}\right)\right\|_{1}\lesssim T^2M^2 \sqrt{\frac{\log (M)+t}{N}},
\end{equation}
with a probability at least $1-8 e^{-c_1 N t}$, which proves Theorem~\ref{thr_1}.
\end{IEEEproof}

\begin{figure*}[t]
	\centering
    \subfloat[]
		{\includegraphics[width=0.6\columnwidth]{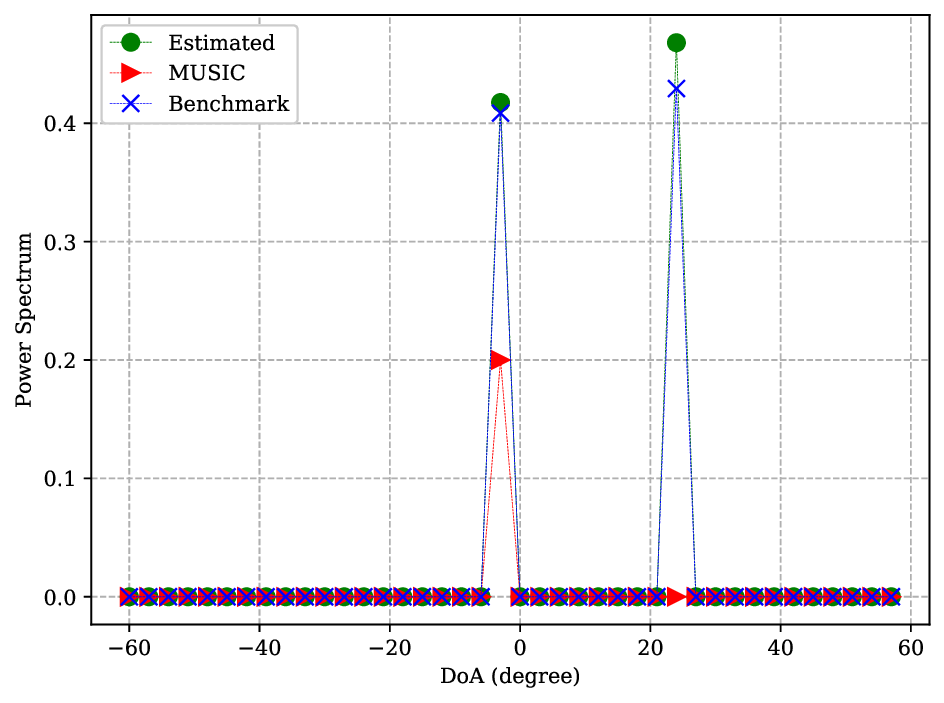}}\qquad 
    \subfloat[]
		{\includegraphics[width=0.6\columnwidth]{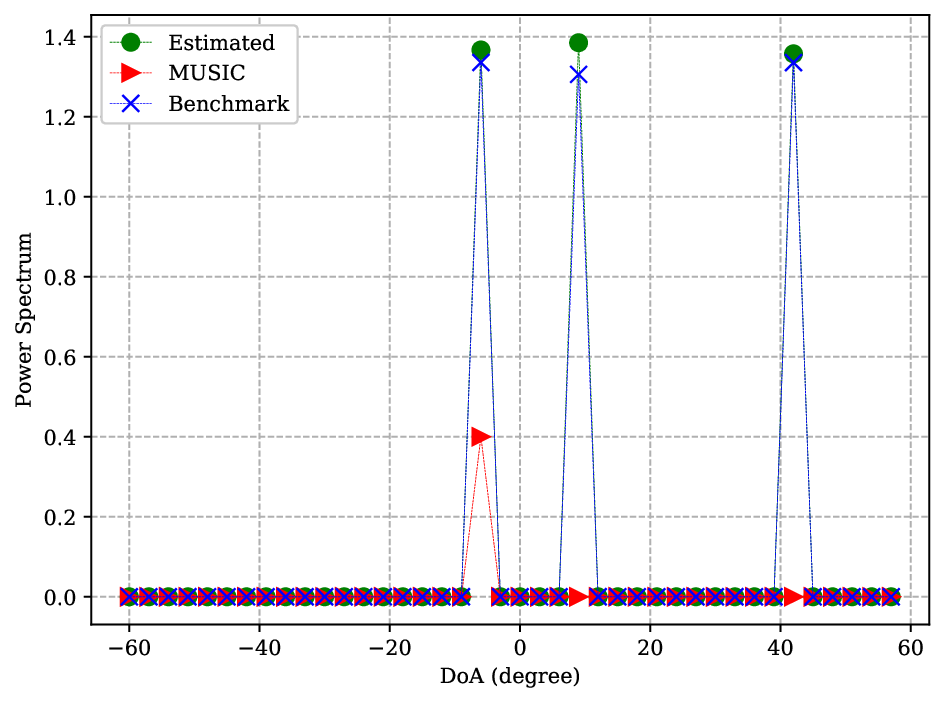}}\qquad
    \subfloat[]
		{\includegraphics[width=0.6\columnwidth]{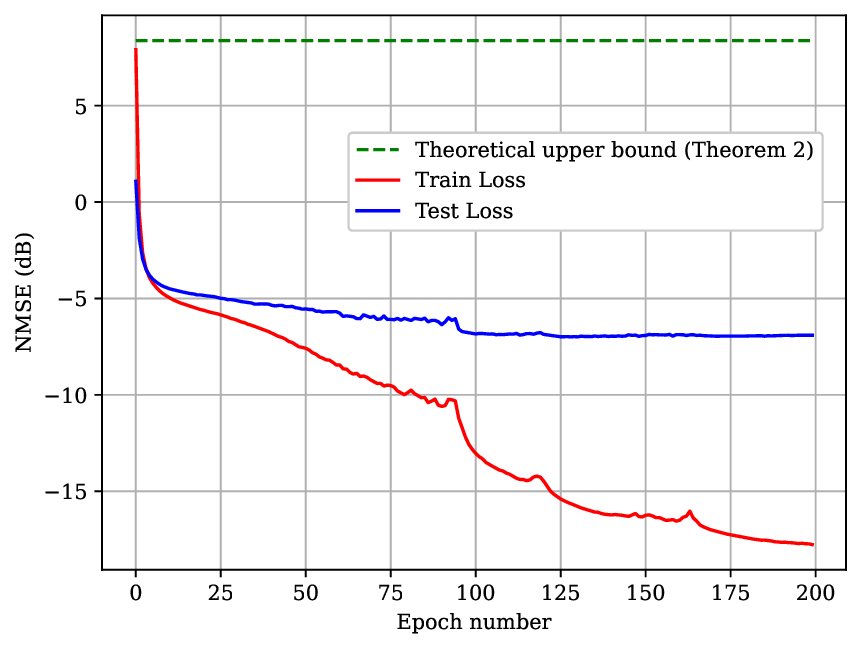}} \vspace{-5pt}
	\caption{  The estimated DoAs by LISTA from one-bit data for (a) two targets, and (b) three targets in comparison with the MUSIC approach. (c) Convergence of LISTA with respect to the number of epochs for the task of DoA estimation from one-bit data. As observed, LISTA  converges even under  coarse quantization of high-resolution measurements.}
\vspace{-5pt}
\label{fig::2}
\end{figure*}
\section{Numerical Illustrations}\label{sec:4}
We exploit several experiments to demonstrate the effectiveness and performance of the deep unrolled network for DoA estimation from one-bit data. While our formulations can be extended to any arrangement of arrays, we used a uniform linear array with $d=\lambda/2$ spacing and $M=\{8,16\}$ antennas equipped with one-bit ADCs. In all of our experiments, the DoA interval is set to $\Delta \theta=1^{\circ}$, with $K=2$ and $K=3$ targets are assumed to be in random locations in the interval $[-60^{\circ},60^{\circ}]$. To generate the one-bit dither data for training, we first generate  
$N=10^{4}$ snapshots of the received signal according to~\eqref{eq2}. Then, we generate the one-bit data by considering the uniform dithering scheme according to~\eqref{one-bit}. Subsequently, the estimated covariance matrix from one-bit data is computed as in~\eqref{cov_est}. In our experiments, we generated $S=1600$ data samples for training and $400$ for validation. Fig.~\ref{fig::2}(a) and Fig.~\ref{fig::2}(b) illustrate the performance of LISTA in estimating the target DoAs for scenarios with $K=2$ targets and $M=8$ one-bit antennas, and $K=3$ targets with $M=16$ one-bit antennas. As observed, LISTA can successfully detect the locations of two and three targets, whereas the MUSIC algorithm can only detect one target. This demonstrates the capability of LISTA to adapt to severely distorted measurements. Fig.~\ref{fig::2}(c) depicts the empirical training and validation losses in comparison with the theoretical convergence bound presented in Theorem~\ref{thr_1}. In this plot, we considered $K=2$ targets and $M=8$ one-bit antennas. To plot the theoretical convergence bound as expressed in \eqref{bound}, we assigned $s=2$, $B=\frac{1}{12}$, $c_2=0.1$, $t=0.01$, $C=2$ and $I=10$. Moreover, one can observe robustness of LISTA to coarse quantization of measurements in terms of convergence of training and validation losses.

\section{Discussion}
\label{sec:5}
In this paper, we explored the application of deep unrolling in addressing the DoA estimation problem under one-bit quantization of measurements. A compressed sensing formulation was presented for this task and  subsequently solved by training  LISTA under low-resolution data. We also provided the convergence guarantee for  our  proposed methodology. Our Numerical results verify the  effectiveness of our approach in comparison with  the well-known MUSIC algorithm.

\bibliographystyle{IEEEtran}
\bibliography{references}

\end{document}